\documentclass[11pt,a4]{article}
\pdfoutput=1

\usepackage{amsfonts}
\usepackage{amsmath}
\usepackage{amssymb}
\usepackage{mathrsfs}
\usepackage[english]{babel}
\usepackage[T1]{fontenc}
\usepackage[latin1]{inputenc}
\usepackage{epsfig}
\usepackage{textcomp}
\usepackage{graphicx}
\usepackage{epstopdf}
\usepackage{subfigure}

\numberwithin{equation}{section}
\def\a{\alpha}
\def\b{\beta}
\def\g{\gamma}

\def\d{\delta}

\def\z{\zeta}

\def\la{\lambda}

\def\m{\mu}
\def\n{\nu}

\def\r{\rho}

\def\o{\omega}

\providecommand{\abs}[1]{\lvert#1\rvert}

\providecommand{\bbabs}[1]{\bigg\lvert#1\bigg\rvert}

\newcommand{\p}{\prime}
\newcommand{\beq}{\begin{equation}}
\newcommand{\eeq}{\end{equation}}

\begin{document}

\date{\mbox{}}

\title{
{\bf \huge Pressure ef\mbox{}fects in the weak-f\mbox{}ield limit of \boldmath $f(R) = R + \alpha R^2$ gravity}
 \\[8mm]
}

\author{
Fulvio Sbis\`a$^{1}$\thanks{fulviosbisa@gmail.com} \hspace{1pt}, Oliver F.\ Piattella$^{1, 2}$\thanks{oliver.piattella@cosmo-ufes.org} \hspace{3pt} and Sergio E.\ Jor\'as$^{3}$\thanks{joras@if.ufrj.br}
\\[8mm]
\normalsize\it
$^1$ Departamento de F\'isica, Universidade Federal do Esp\'irito Santo,\\
\normalsize\it
Avenida Fernando Ferrari, 514, CEP 29075-910, Vit\'oria, ES, Brazil \vspace{.5cm} \\
\normalsize\it
$^2\,$ Institut f\"ur Theoretische Physik, Ruprecht-Karls-Universit\"at Heidelberg,\\
\normalsize\it
Philosophenweg 16, 69120 Heildelberg, Germany \vspace{.5cm} \\
\normalsize\it
$^3\,$ Instituto de F\'isica, Universidade Federal do Rio de Janeiro,\\
\normalsize\it
C.P.\ 68528, CEP 21941-972, Rio de Janeiro, RJ, Brazil \\
}

\maketitle

\setcounter{page}{1}
\thispagestyle{empty}

\begin{abstract}
We investigate the linear regime of $f(R) = R + \alpha R^2$ gravity for static, spherically symmetric and asymptotically f\mbox{}lat conf\mbox{}igurations of matter. We show that, in vacuum and deep inside the range of the extra scalar degree of freedom, the post-Newtonian parameter $\gamma$ is not equal to 1/2, as established in the literature, but it assumes larger values depending on the pressure of the star. We provide an explicit expression for $\g$ in terms of the mass, of the integrated pressure of the star and of the ratio between the star's radius and the range of the extra degree of freedom. We corroborate our results by providing numerical solutions for the case of a neutron star.
\end{abstract}

\smallskip

\section{Introduction}
\label{Introduction}

Among the numerous extensions of the theory of General Relativity (GR), in the past two decades considerable attention has been given to $f(R)$ gravity (where $R$ is the Ricci scalar) as a possible alternative to dark energy, and to explore modif\mbox{}ications of gravity in the strong gravity regime. The action describing this modif\mbox{}ication of GR is obtained from the Einstein-Hilbert action by simply trading $R$ for a generic non-linear function of it \cite{Sotiriou:2008rp, DeFelice:2010aj, Nojiri:2010wj, Nojiri:2017ncd}, that is:
\begin{equation}
	S = \frac{c^4}{16 \pi G_{_{\! N}}} \int \! d^{4} x \, \sqrt{-g} \, f(R) + S_{_{\! M}} \; ,
\end{equation}
where $G_{_{\! N}}$ is Newton's constant and $S_{_{\! M}}$ is the action for the matter f\mbox{}ields.

In the metric formalism of $f(R)$ \cite{Buchdahl1970}, which we adopt here, this apparently innocent procedure spoils the feature of GR of having f\mbox{}ield equations of second-order. In metric $f(R)$ the evolution is in fact described by fourth-order dif\mbox{}ferential equations, which is in general worrying from the point of view of Ostrogradsky instability \cite{Ostrogradsky:1850fid}. It can be however shown that one of the assumptions of Ostrogradsky's theorem (non-degeneracy) is violated \cite{Woodard:2006nt}, and that the theory is safe. A generic prescription on the form of $f$, at least for small modif\mbox{}ications of GR, is that $f_{_{\!RR}} > 0\,$, in order to avoid the Dolgov-Kawasaki instability \cite{Dolgov:2003px}.\footnote{We indicate derivatives with respect to $R$ with a subscript $\!\!\!\phantom{f}_{_{\!R}}$.}

A well-known property of $R$-regular $f(R)$ theories is that they can be mapped into a scalar-tensor theory (in fact, a Brans-Dicke theory with parameter $\o = 0$ and a non-trivial potential), where the derivative $f_{_{\!R}}$ corresponds to the scalar component.\footnote{A $f(R)$ theory is called $R$-regular if the second derivative $f_{_{\!RR}}$ never vanishes.} In comparison to GR, this component therefore acts ef\mbox{}fectively as an extra degree of freedom. As usual, it is possible to pass from the Jordan to the Einstein frame by means of a conformal transformation. The coupling of the scalar degree of freedom to the matter sector, which is induced by the transformation, is at the basis of the chameleon mechanism \cite{Khoury:2003aq, Khoury:2003rn}.

In this paper we focus on the linear regime of $f(R)$ gravity in a static, spherically symmetric and asymptotically f\mbox{}lat conf\mbox{}iguration, providing a simple model for a star. This regime has been extensively considered, although at times with cosmological asymptotics, especially in connection with Solar System tests (see e.g.\ \cite{Accioly:1998nm, Accioly:2000gd, Dick:2003dw, Nojiri:2003ft, Chiba:2003ir, Soussa:2003re, Rajaraman:2003st, Easson:2004fq, Olmo:2005zr, Olmo:2005hc, Navarro:2005gh, Cembranos:2005fi, Capozziello:2005bu, Clifton:2005aj, Barrow:2005dn, Shao:2005wt, Navarro:2005da, Capozziello:2006jj, Multamaki:2006zb, Jin:2006if, Ruggiero:2006qv, Navarro:2006mw, Baghram:2007df, Zhang:2007ne, Capozziello:2007wc, Hu:2007nk, Henttunen:2007bz, Nojiri:2007as, Capozziello:2007ms, Capozziello:2007id, Multamaki:2007jk, Capozziello:2007eu, Cognola:2007zu, Capozziello:2009vr, Capozziello:2010iha, Capozziello:2010wt, Berry:2011pb, Chiba:2006jp, Olmo:2006eh, Kainulainen:2007bt, Guo:2013fda}). A well-established conclusion in this regard is that, outside the star and well within the range of the extra scalar degree of freedom, the metric potentials display the Newtonian $1/r$ behavior but have dif\mbox{}ferent magnitude. This dif\mbox{}ference is encoded in the post-Newtonian parameter (PPN) $\gamma\,$, whose value is found to be equal to 1/2 and therefore in clear disagreement with observation (which conf\mbox{}irms the GR value $\gamma = 1$ within few parts in $10^5$ \cite{Will:2005va}). An analysis of the behaviour of the PPN parameter $\g$ in non-minimally coupled models of gravity (which have a $f(R)$ limit) can be found in \cite{Bertolami:2013qaa}. The behaviour of the post-Newtonian parameters $\b$ and $\g$ in a generic scalar-tensor theory, although with very crude assumptions on the structure of the star, is studied in \cite{Hohmann:2013rba, Hohmann:2017qje}.

The $\g = 1/2$ result is based on the assumption of negligible pressure of the spherically symmetric energy-momentum conf\mbox{}iguration. Here we show that $\gamma$ is in general larger than 1/2 when the pressure is taken into account. Our analysis is limited to the model $f(R) = R + \alpha R^2$, also known as Starobinsky model \cite{Starobinsky:1980te}, which is one of the most successful models in inf\mbox{}lationary cosmology \cite{Martin:2013nzq}. Nevertheless, we expect our analysis to be relevant for any $f(R)$ theory such that $f$ is analytic in $R = 0\,$. We also comment on a relativistic limit, i.e.\ when the spherical conf\mbox{}iguration of energy-momentum has a relativistic equation of state, in which case the $\gamma = 1$ limit is recovered. Furthermore, we explore the role of mild non-linear ef\mbox{}fects on the behaviour of gravity.

The paper is organized as follows: in section \ref{Field equations} we set up our notation and derive the equations of motions for our system. In section \ref{Linearization} we implement the linear approximation of the latter equations and derive the external and internal solutions. In section \ref{Linear gravity} we discuss the consistency of our approximation and draw some conclusions about the behaviour of gravity. In section \ref{Numerical solutions} we support our analysis by solving numerically the non-linear equations of motion. We present our conclusions and comments in section \ref{Conclusions}.

We adopt the ``mostly plus'' signature $(-,+,+,+)$ and, unless stated otherwise, we use units of measure where $c = 1\,$.

\section{Field equations}
\label{Field equations}

We consider the $f(R)$ gravity model given by:
\begin{equation}
	f(R) = R + \alpha R^2 \; ,
\end{equation}
in the metric formalism. The f\mbox{}ield equations resulting from the variation of the action with respect to the metric are the following:
\begin{equation}\label{metric eq}
	\big( 1 + 2 \a R \big) \, R_{\m\n} - \frac{1}{2} \, g_{\m\n} \, \big( R + \alpha R^2 \big) - 2 \a \, \big( \nabla_{\!\m} \nabla_{\!\n} - g_{\m\n} \, \Box \big) \, R = 8 \pi G_{_{\! N}} \, T_{\m\n} \; ,
\end{equation}
where the connection is the Levi-Civita one, so the Ricci scalar is a functional of the metric $R = R \, [g_{\m\n}]$, $\Box = g^{\m\n} \, \nabla_{\!\m} \, \nabla_{\!\n}$ is the (curved space) d'Alembert operator and
\begin{equation}
	T_{\m\n} = - \frac{2} {\sqrt{-g}} \frac{\d S_{_{\! M}}}{\d g^{\m\n}} \; .
\end{equation}

Def\mbox{}ining $\z = (f_{_{\!R}} - 1)/2$, the fourth-order equation (\ref{metric eq}) can be shown to be equivalent to the second-order system for the metric $g_{\m\n}$ and the scalar f\mbox{}ield $\z$ \cite{Teyssandier:1983zz}:
\begin{align}
	\big( 1 + 2 \z \big) \, G^\mu{}_\nu &= - 3 m^2 \, \z^2 \, \delta^\mu{}_\nu + 2 \, \Big( g^{\mu\la} \nabla_\nu \, \partial_\la - \d^\mu{}_\nu \, \Box \Big) \z + 8 \pi G_{_{\! N}} \, T^\mu{}_\nu \; , \label{metric eq gen} \\[3mm]
	\Box \, \z - m^2 \z &= \frac{4 \pi G_{_{\! N}}}{3} \, T \; , \label{chi eq gen}
\end{align}
where $T$ is the trace of the energy-momentum tensor and we def\mbox{}ined:
\begin{equation}
	m^2 \equiv \frac{1}{6\alpha}
\end{equation}
as the mass associated to the scalar degree of freedom. The metric and the scalar f\mbox{}ield are independent degrees of freedom, as far as the initial value problem of the system (\ref{metric eq gen})--(\ref{chi eq gen}) is concerned, however when the solutions of (\ref{metric eq gen})--(\ref{chi eq gen}) are considered the scalar f\mbox{}ield is related to the scalar curvature by $\z = \a R\,$.

\subsection{Static and spherically symmetric case}

As a model for a non-rotating star, we assume a static, spherically symmetric metric:
\begin{equation}
	ds^2 = - A(r) \, dt^2 + B(r) \, dr^2 + r^2 d\theta^2 + r^2 \sin^2 \!\theta \, d\phi^2 \; , 
\end{equation}
and a perfect f\mbox{}luid-type energy-momentum tensor:
\begin{equation}
	T^0{}_0 = -\rho(r)\;, \qquad T^r{}_r = T^\theta{}_\theta = T^\phi{}_\phi = p(r) \; , 
\end{equation}
where $\rho$ and $p$ are the density and pressure of the star, respectively, and they depend only on the radial coordinate.

The f\mbox{}ield equations can be straightforwardly computed and read:
\begin{subequations}\label{systemTOV}
\begin{align}
	&\Big( 1 + 2 \z + r \z' \Big) \, \frac{1}{Br} \frac{A'}{A} = \frac{1 + 2 \z}{r^2} \, \bigg(1 - \frac{1}{B} \bigg) - 3 m^2 \z^{2} - \frac{4}{B r} \, \z' + 8 \pi G_{_{\! N}} \, p \; , \label{Aeqfull} \\[5mm]
	&\Big( 1 + 2 \z + r \z' \Big) \, \frac{1}{Br} \frac{B'}{B} = \frac{1 + 2 \z}{r^2} \, \bigg( \frac{1}{B} - 1 \bigg) + 3 m^2 \z^{2} + \frac{2}{B} \, \bigg( \z'' + \frac{2}{r} \, \z' \bigg) + 8 \pi G_{_{\! N}} \, \rho \; , \label{Beqfull} \\[5mm]
	&\frac{1}{B} \, \bigg[ \, \z'' + \left( \frac{2}{r} + \frac{A'}{2A} - \frac{B'}{2B} \right) \, \z' \, \bigg] = m^{2} \z + \frac{4 \pi G_{_{\! N}}}{3} \, \big( 3p - \r \big) \; , \label{chieqfull}
\end{align}
\end{subequations}
where the prime denotes derivation with respect to $r$. Using eqs.~\eqref{Aeqfull} and \eqref{chieqfull}, the equation \eqref{Beqfull} can be cast in the form:
\begin{multline}
	\frac{1 + 2 \z}{r} \, \frac{B'}{B^2} = \left(1 + \frac{r \z'}{1 + 2 \z + r \z'}\right) \, \left[ \, \frac{1 + 2 \z}{r^2} \left( \frac{1}{B} - 1 \right) + 3 m^{2} \z^{2} \, \right] + 2 m^{2} \z + \\[1mm]
	+\frac{4}{B} \, \frac{{\z'}^2}{1 + 2 \z + r \z'} + 8 \pi G_{_{\! N}} \left( \frac{1 + 2 \z}{1 + 2 \z + r \z'} \, p + \frac{2}{3} \, \rho \right) \; . \label{Beqter}
\end{multline}

\section{Linearization of the f\mbox{}ield equations}
\label{Linearization}

We now want to study analytically the above system of equations, in the linear regime. To this aim it is useful to introduce the gravitational potentials $\Phi$ and $\Psi$:
\begin{equation}
	A(r) = 1 + 2 \, \Phi(r) \; , \qquad B^{-1}(r) = 1 + 2 \, \Psi(r) \; .
\end{equation}
Since we consider asymptotically f\mbox{}lat solutions, we look for (approximated) solutions of the equations above such that $\z\,$, $\Phi$ and $\Psi$ decay to zero when $r \to \infty\,$.

\subsection{The scalar degree of freedom}

Assuming that the following conditions hold:
\begin{align}
	\abs{\Phi} &\ll 1 \; , & \abs{\Psi} &\ll 1 \; , & r \, |\Phi' + \Psi'| \ll 1 \; , \label{wfcond1}
\end{align}
the evolution equation for the scalar degree of freedom, eq.~\eqref{chieqfull}, can be greatly simplif\mbox{}ied as follows:
\begin{equation}\label{chieq}	
       \z'' + \frac{2}{r} \, \z' - m^2 \z = \frac{4 \pi G_{_{\! N}}}{3} \, \big( 3 p - \rho \big) \; ,
\end{equation}
and can be readily solved, since it does not contain the gravitational potentials. The (asymptotically decaying) Green's function associated to the dif\mbox{}ferential operator acting on $\z$ is the following:
\begin{equation}
	G(\mathbf r,\mathbf r') = -\frac{1}{4\pi |\mathbf r - \mathbf r' |}e^{-m |\mathbf r - \mathbf r' |} \; ,
\end{equation}
so that the complete solution of eq.~\eqref{chieq} which decays at inf\mbox{}inity reads:
\begin{equation}\label{alphachisolutionfull}
	\z (\mathbf r) = \frac{G_{_{\! N}}}{3} \int_{V_\star} \frac{e^{-m |\mathbf r - \mathbf r' |}}{|\mathbf r - \mathbf r'|} \, \Big(\rho (\mathbf r') - 3p(\mathbf r') \Big) \, d^3\mathbf r' \; ,
\end{equation}
where the integration is performed on $V_\star$, the spherical volume occupied by the star.

Since $\rho$ and $p$ depend only on the radial coordinate, the angular integration in (\ref{alphachisolutionfull}) can be performed exactly. This is achieved by making explicit the modulus:
\begin{equation}
	|\mathbf r - \mathbf r'| = \sqrt{r^2 + {r'}^2 - 2rr' \cos \b} \; ,
\end{equation}
where $\b$ is the angle between the directions of $\mathbf r$ and $\mathbf r'$, and then using the integral:
\begin{equation}
	\int_{-1}^1 d(\cos\b) \, \frac{e^{-m\sqrt{r^2 + {r'}^2 - 2rr'\cos\b}}}{\sqrt{r^2 + {r'}^2 - 2rr'\cos\b}} = \frac{e^{-m|r - r'|} - e^{-m(r + r')}}{mrr'} \; . \label{integral}
\end{equation}

\subsubsection{External and internal solutions}
\label{External and internal solutions}

Let us consider the region outside the star. The integration of eq. (\ref{alphachisolutionfull}) then gives:
\begin{equation}\label{alphachisolutionmatter}
	\z_{\rm ext}(r) = \frac{G_{_{\! N}}}{3} \left( \tilde{M}_{\star} - 3 \tilde{P}_{\star} \right) \frac{e^{-mr}}{r} \; ,
\end{equation}
where we def\mbox{}ined:
\begin{align}\label{MCandPtildedefinitions}
	\tilde{M}_{\star} &\equiv 4 \pi \int_{0}^{r_\star} \frac{\sinh mr}{mr} \, \rho \, r^2 dr \; , & \tilde{P}_{\star} &\equiv 4 \pi \int_{0}^{r_\star} \frac{\sinh mr}{mr} \, p \, r^2 dr \; ,
\end{align}
and indicated with $r_\star$ the radius of the star. The quantities $\tilde{M}_{\star}$ and $\tilde{P}_{\star}$ are related respectively to the integrated density and pressure of the star, and in fact reduce to them when $m r_\star \ll 1\,$:
\begin{align}\label{MCandPdefinitionsmrsmall}
	\tilde{M}_{\star} &\xrightarrow[m r_\star \ll 1]{} M_{\star} \equiv 4 \pi \int_{0}^{r_\star} \!\! \rho \, r^2 dr \; , & \tilde{P}_{\star} &\xrightarrow[m r_\star \ll 1]{} P_{\star} \equiv 4 \pi \int_{0}^{r_\star} \!\! p \, r^2 dr \; .
\end{align}
In this case, which corresponds to the range of the scalar degree of freedom being much larger than the radius of the star, outside of the star there exists a region where the exterior solution becomes:\footnote{This result is equivalent to eq.~(13) of \cite{Chiba:2006jp} specialised for the $R + \alpha R^2$ model, apart from the fact that here the contribution due to the pressure is taken into account.}
\begin{equation}\label{chisolutionmattermrsmall}
	r_{\star} \leq r \ll m^{-1} \qquad \Rightarrow \qquad \z_{\rm ext}(r) = \frac{G_{_{\! N}}}{3 r}\, \big( M_{\star} - 3 P_{\star} \big) \; .
\end{equation}
It is suggestive to introduce a characteristic radius associated with the extra scalar degree of freedom:
\begin{equation}\label{rchi}
	r_{\!\z} \equiv 2 G_{_{\! N}} \big( \tilde{M}_{\star} - 3 \tilde{P}_{\star} \big) \; ,
\end{equation}
which may be considered as the ``Schwarzschild radius'' of $\z\,$. In terms of $r_{\!\z}$ the external solution (\ref{alphachisolutionmatter}) reads
\begin{equation}\label{alphachisolutionexternal}
	\z_{\rm ext}(r) = \frac{r_{\!\z}}{6 r} \, e^{-mr} \; .
\end{equation}

Let us now consider the behaviour of $\z$ inside the star. Because of the absolute value $\abs{r - r'}$ appearing in (\ref{integral}), it is convenient to split the radial integration into two parts. We then obtain
\begin{equation}\label{alphachiapproxfull}
	\z_{\rm int}(r) = \frac{G_{_{\! N}}}{3} \, \bigg[ \, \frac{e^{-mr}}{r} \, \Big( \tilde{M}(r) - 3 \tilde{P}(r) \Big) + \frac{\sinh mr}{mr} \, \tilde{S}(r) \, \bigg] \; ,
\end{equation}
where we def\mbox{}ined:
\begin{align}\label{tildeMrho e tildeP}
	\tilde{M}(r) &\equiv 4 \pi \int_{0}^{r} \frac{\sinh my}{my} \, \rho \, y^2 dy \; , & \tilde{P} (r) &\equiv 4 \pi \int_{0}^{r} \frac{\sinh my}{my} \, p \, y^2 dy \; ,
\end{align}
and:
\begin{equation}\label{tildeS}
	\tilde S(r) \equiv 4 \pi \int_{r}^{r_\star} \!\! e^{-my} \, \Big( \rho(y) - 3p(y) \Big) \, y \, dy \; .
\end{equation}
Note that
\begin{align}
	\tilde{M}(r_{\star}) &= \tilde{M}_{\star} \; , & \tilde{P} (r_{\star}) &= \tilde{P}_{\star} \; , & \tilde{S} (r_{\star}) &= 0 \; ,
\end{align}
so the expression (\ref{alphachiapproxfull}) actually holds also outside the star provided we def\mbox{}ine $\tilde{M}(r)\,$, $\tilde{P}(r)$ and $\tilde{S}(r)$ to be continuous, and constant outside the star. Furthermore, the limit of (\ref{alphachiapproxfull}) for $r \to 0$ is well-def\mbox{}ined, since $\tilde{M}(r)/r \to 0$ and $\tilde{P}(r)/r \to 0\,$, and we have:
\begin{equation}\label{alphachiapproxorigin}
	\z_{\rm int}(0) = \frac{G_{_{\! N}}}{3} \, \tilde{S}(0) \; .
\end{equation}

A word on the case of relativistic matter. It is apparent that when $p = \rho/3$ the functions $\tilde{M}(r) - 3 \tilde{P}(r)$ and $\tilde{S}(r)$ vanish identically, and therefore the scalar degree of freedom vanishes identically as well (both inside and outside the star). In this case, therefore, the scalar degree of freedom is ``masked''. We further comment on this point in section \ref{Relativistic matter}.

\subsection{The gravitational potentials}

Let us consider the extra conditions
\begin{align}
	\abs{\z} &\ll 1 \; , & r \, \abs{\z'} &\ll 1 \; , & r \, m^{2} \z^2 &\ll \abs{\z'} \; , \label{wfcond3}
\end{align}
and suppose that the term containing ${\z'}^2$ can be neglected in eq.~(\ref{Beqter}). All these assumptions are to be discussed more in detail below. Under this approximation the f\mbox{}ield equations \eqref{Aeqfull} and \eqref{Beqter} simplify to:
\begin{subequations}\label{systemTOVlin}
\begin{align}
\label{Phieq}	r \Phi' &= - \Psi - 2 \, r \z' + 4 \pi G_{_{\! N}} \, r^2 p \; , \\[2mm]
\label{Psieq}	\big( r \Psi \big)' &= - r^2 m^{2} \z - \frac{4 \pi G_{_{\! N}} }{3} \, r^2 \big( 2\rho + 3p \big) \; .
\end{align}
\end{subequations}
Knowing the explicit solution for $\z$, we can straightforwardly obtain those for the gravitational potentials $\Psi$ and $\Phi$.

\subsubsection{The potential $\Psi$}

Let us start considering the internal solution for the potential $\Psi\,$. Integrating eq.~\eqref{Psieq} from $r = 0$ to a certain radius $r \leq r_{\star}$ we obtain:
\begin{equation}
\label{Psisollinearfull}
	r \Psi(r) = - m^{2} \!\int_0^r \! \z(y) \, y^2 dy - \frac{G_{_{\! N}}}{3} \, \Big( 2 M(r) + 3P(r) \Big) \; ,
\end{equation}
where we assumed that $r \Psi(r) \to 0$ for $r\to 0$ (which should be safe, since we expect the gravitational f\mbox{}ield to vanish at the center of the star). Let us now def\mbox{}ine:
\begin{equation}\label{Xidefinition}
	\Xi(r) \equiv \frac{m^{2}}{G_{_{\! N}}} \int_0^{r} \! \z(y) \, y^2 dy \; ,
\end{equation}
and adopt the notation:\footnote{Note that $\Xi_{\star}$ is the amount of scalar curvature enclosed by the star, divided by $6 G_{_{\! N}}\,$.}
\begin{equation}\label{Xiconstant}
	\Xi_{\star} \equiv \Xi(r_\star) \; .
\end{equation}
Once the conf\mbox{}iguration of $\r$ and $p$ is known, the function $\Xi(r)$ can be computed (in principle exactly, but most likely performing the integrations numerically) by employing eqs.~\eqref{alphachiapproxfull}--\eqref{tildeS}. The solution for $\Psi$ inside the star can thus be written as:
\begin{equation}\label{Psisolinternal}
	\Psi_{\rm int}(r) = - \frac{G_{_{\! N}}}{r} \, \bigg( \frac{2}{3} M(r) + P(r) + \Xi(r) \bigg) \; .
\end{equation}

The external ($r \ge r_\star$) solution for $\Psi$ can be obtained by f\mbox{}inding a primitive of the right hand side of (\ref{Psieq}) in vacuum (since we know explicitly the external solution \eqref{alphachisolutionmatter} for $\z$) and matching the resulting function with the internal solution (\ref{Psisolinternal}) at the star's surface. We get:
\begin{equation}
	r\Psi_{\rm ext}(r) = \frac{r_{\!\z}}{6} \, \Big[ e^{-mr} \big( 1 + mr \big) - e^{-mr_\star} \big( 1 + m r_\star \big) \Big] - G_{_{\! N}}  \bigg( \frac{2}{3} M_{\star} + P_{\star} + \Xi_{\star} \bigg) \; .
\end{equation}
When $r \to \infty$, the right hand side of the above equation tends to a f\mbox{}inite value $C$:
\begin{equation}\label{C2formula}
	C \equiv - G_{_{\! N}} \Bigg[ \bigg( \frac{\tilde{M}_{\star}}{3} - \tilde{P}_{\star} \bigg) \, e^{-mr_\star} \, \big( 1 + mr_\star \big) + \frac{2 M_{\star}}{3} + P_{\star} + \Xi_{\star} \, \Bigg] \; ,
\end{equation}
which def\mbox{}ines a geometric mass $M_g = - C/G_{_{\! N}}$ distinct from $M_{\star}$ or, equivalently, a new characteristic radius $r_{\!g} \equiv - 2C$ for the gravitational potential. We can then write in a more compact form the external solution for $\Psi(r)$:
\beq \label{Psisolexternal1}
	\Psi_{\rm ext}(r) = \big( 1 + mr \big) \, \frac{r_{\!\z}}{6 r} \, e^{-mr} - \frac{r_{\!g}}{2r} \; ,
\eeq
from which it is evident that the external solution correctly decays to zero asymptotically.

\subsubsection{The potential $\Phi$}

Let us now consider the gravitational potential $\Phi$. Regarding the external solution, also in this case it can be found by looking for a primitive of the equation (\ref{Phieq}) in vacuum (using the explicit external solutions \eqref{alphachisolutionmatter} and \eqref{Psisolexternal1} respectively for $\z$ and $\Psi$), and imposing it to decay to zero asymptotically. We obtain:
\beq \label{Phisolutionexternal1}
	\Phi_{\rm ext}(r) = -\frac{r_{\!\z}}{6r} \, e^{-mr} - \frac{r_{\!g}}{2r} \; .
\eeq

The internal solution can be expressed in terms of $\Psi\,$, $\z$ and $p$ (and so, ultimately, in terms of $\r$ and $p$) by integrating eq.~\eqref{Phieq} and using eqs.~\eqref{alphachiapproxfull} and \eqref{Psisolinternal}. We get:
\begin{equation}\label{Phisolfulllinear}
	\Phi_{\rm int}(r) = \Phi(0) - \int_0^r\frac{\Psi(y)}{y} \, dy - 2 \z(r) + \frac{2 G_{_{\! N}}}{3} \, \tilde{S}(0) + 4 \pi G_{_{\! N}} \int_0^{r} \!p(y) \, y \, dy \; ,
\end{equation}
where the integration constant $\Phi(0)$ has to be chosen so as to match the external solution (\ref{Phisolutionexternal1}) at the star's surface. Again, although $\Phi_{\rm int}$ is expressed exactly in terms of (multiple integrations of) $\r$ and $p$, in practice it will be computed by performing the integrations numerically.

\section{Gravity in the linear regime}
\label{Linear gravity}

\subsection{Consistency of the approximation}
\label{Consistency approximation}

Let us comment now on the meaning and consistency of the assumptions (\ref{wfcond1}) and (\ref{wfcond3}), which we may group as follows
\begin{align}
	\abs{\Phi} &\ll 1 \; , & \abs{\Psi} &\ll 1 \; , & \abs{\z} &\ll 1 \; , \label{first group} \\[5mm]
	r \, |\Phi' + \Psi'| &\ll 1 \: , & r \, \abs{\z'} &\ll 1 \; , & r \, m^{2} \z^2 &\ll \abs{\z'} \; , \label{second group}
\end{align}
and on the assumption that the term containing ${\z'}^2$ can be neglected in eq.~(\ref{Beqter}).

\subsubsection{External solutions}

Let us consider the external solutions. From the expressions (\ref{alphachisolutionexternal}), (\ref{Psisolexternal1}) and (\ref{Phisolutionexternal1}) it is apparent that the f\mbox{}irst group of assumptions (\ref{first group}) is essentially equivalent (unless $m r_{\star}$ is much bigger than one) to the condition of the characteristic radii $r_{\!\z}$ and $r_{\!g}$ being much smaller than the star's radius
\begin{align}
	r_{\!\z} &\ll r_{\star} \; , & r_{\!g} &\ll r_{\star} \; .
\end{align}
It is not dif\mbox{}f\mbox{}icult to see that, in this case, the assumptions (\ref{second group}) are automatically satisf\mbox{}ied. Note in fact that
\begin{align*}
	r \, |\Phi' + \Psi'| &= \bbabs{\frac{r_{\!\z}}{6 r} \, f_{1}(mr) - \frac{r_{\!g}}{r}} \; ,\quad \, & r \, \abs{\z'} &= \frac{r_{\!\z}}{6 r} \, f_{2}(mr) \; ,\quad \, & \frac{r \, m^{2} \z^2}{\abs{\z'}} &= \frac{r_{\!\z}}{6 r} \, f_{3}(mr) \; ,
\end{align*}
where we introduced the functions
\begin{align*}
	f_{1}(x) &= x^{2} e^{-x} \; , & f_{2}(x) &= (1 + x) \, e^{-x} \; , & f_{3}(x) &= \frac{x^{2}}{1 + x} \, e^{-x} \; .
\end{align*}
Since $f_{1}$, $f_{2}$ and $f_{3}$ are smaller than $1$ on $[\, 0 , + \infty)\,$, the thesis follows.

When $m r_{\!\star} \gg 1\,$, instead, the condition $r_{\!\z} \ll r_{\star}$ becomes superf\mbox{}luous, since the exponential $e^{-mr}$ supresses all the terms containing $r_{\!\z}\,$. Therefore, in this case the assumptions (\ref{first group}) and (\ref{second group}) are equivalent to the condition $r_{\!g} \ll r_{\!\star}$ only. This is consistent with GR being recovered in the $m \to \infty$ limit, since in GR the non-linear terms in the equations of motion are negligible whenever the characteristic radius (i.e.\ the Schwarzschild one) is much smaller than the star's radius.

Regarding the role of the ${\z'}^2$ term in eq.~(\ref{Beqter}), let us consider the latter equation imposing the conditions (\ref{first group}) and (\ref{second group}) but without neglecting ${\z'}^2$. Instead of (\ref{Psieq}), we obtain:
\beq
	\big( r \Psi \big)' = - r^{2} m^{2} \z - 2 \, r^{2} {\z'}^2 - \frac{4 \pi G_{_{\! N}}}{3} \, r^{2} \big( 2 \rho + 3 p \big) \; ,
\eeq
which can be readily integrated (in vacuum) to give:
\beq \label{Psi ext zeta p squared}
	\Psi_{\rm ext}(r) = \big( 1 + mr \big) \, \frac{r_{\!\z}}{6 r} \, e^{-mr} - \frac{r_{\!g}}{2r} + \big( 2 + mr \big) \bigg( \frac{r_{\!\z}}{6 r} \, e^{-mr} \! \bigg)^{\!\!2} \; .
\eeq
By comparing (\ref{Psi ext zeta p squared}) and (\ref{Psisolexternal1}), it is apparent that the term in (\ref{Beqter}) containing ${\z'}^2$ just produces a second order correction to $\Psi_{\rm ext}$, which can be consistently neglected whenever the conditions (\ref{first group}) hold.

We conclude that, as far as the external solutions are concerned, only the weak-f\mbox{}ield conditions (\ref{first group}) need to be assumed, since the conditions (\ref{second group}) and the negligibility of the ${\z'}^2$ term follow from them. In particular this implies that the non-linear terms in the equations of motion are negligible whenever the characteristic radii $r_{\!\z}$ and $r_{\!g}$ are much smaller than the star's radius, which nicely generalizes the analogous result which holds in GR.

\subsubsection{Internal solutions}
\label{Consistency internal solutions}

Regarding the internal solutions, the relationship between the assumptions (\ref{first group}) and (\ref{second group}) is less clear. This is so because the internal solutions depend on the \emph{functions} $\r(r)$ and $p(r)$, instead of on the numbers $r_{\!\z}$ and $r_{\!g}\,$, and do so in a fairly complicated way.

On general grounds, 
we expect the scalar curvature $\z/\a$ to increase when we move inwards, towards the center of the star. This expectation can be substantiated as follows. Let us divide $\z_{\rm int}$ into a ``density'' and a ``pressure'' part
\begin{equation}
	\z_{\rm int}(r) = \z_{\rm int}^{\r}(r) - 3 \, \z_{\rm int}^{p}(r) \; ,
\end{equation}
where
\begin{equation}
	\z_{\rm int}^{\r}(r) = \frac{4 \pi G_{_{\! N}}}{3} \, \Bigg( \frac{e^{-mr}}{r} \int_{0}^{r} \frac{\sinh my}{my} \, \rho \, y^2 dy + \frac{\sinh mr}{mr} \int_{r}^{r_\star} \!\! e^{-my} \, \rho(y) \, y \, dy \Bigg) \; ,
\end{equation}
and $\z_{\rm int}^{p}$ is obtained by substituting $\r \to p$ in the expression above. Let us consider the case $m r_{\star} \ll 1\,$. It is then easy to check that, if $\r$ and $p$ are positive, then $\z_{\rm int}^{\r}$ and $\z_{\rm int}^{p}$ are positive and the derivatives ${\z_{\rm int}^{\r}}'$ and ${\z_{\rm int}^{p}}'$ are negative, so it follows that $\abs{\z_{\rm int}}$ is decreasing in the direction of increasing $r$. In the case $m r_{\star} \gg 1\,$, on the other hand, $\z/\a$ is approximately given by the GR relation $\z/\a = R = 8 \pi G_{_{\! N}} (\r - 3 p)$. It is then evident that, if $\r$ and $p$ are positive and decreasing, then $\abs{\z_{\rm int}}$ again is decreasing (in the direction of increasing $r$). Since it is natural to expect $\r$ and $p$ to have the mentioned properties, the analysis of these two limits indeed suggests $\abs{\z_{\rm int}}$ to be decreasing. Were this the case, the assumption $\abs{\z} \ll 1$ (inside \emph{and} outside the star) would be implied by the condition
\beq
\abs{\z_{\rm int}(0)} = \frac{4 \pi G_{_{\! N}}}{3} \int_{0}^{r_\star} \!\! e^{-my} \, \Big( \rho(y) - 3p(y) \Big) \, y \, dy \ll 1 \; .
\eeq

Overall it is likely that, to thoroughly understand the relationship between the assumptions (\ref{first group}) and (\ref{second group}) inside the star, a detailed discussion about the properties of $\r$ and $p$ is needed. Here we prefer not to embark in this discussion, and simply limit our analysis to those conf\mbox{}igurations where (\ref{first group}) and (\ref{second group}) hold, and where the term containing ${\z'}^2$ can be neglected in eq.~(\ref{Beqter}). We check a posteriori, by numerical means, that these assumptions are justif\mbox{}ied for the conf\mbox{}igurations analysed in section \ref{Numerical solutions}.

\subsection{The behaviour of gravity}

We are now in the position of drawing three interesting conclusions.

\subsubsection{Relationship between the masses}
\label{Relationship masses}

The f\mbox{}irst is a relationship between the geometric mass $M_g = - C/G_{_{\! N}}$ and the masses $M_{\star}$ and $\tilde{M}_{\star}$ computed from the density of the star. From eq. (\ref{C2formula}) we have in fact:
\begin{equation}
	M_g = \bigg( \frac{\tilde{M}_{\star}}{3} - \tilde{P}_{\star} \bigg) \, e^{-mr_\star} \big( 1 + m r_\star \big) + \frac{2 M_{\star}}{3} + P_{\star} + \Xi_{\star} \; ,
\end{equation}
so the geometrical mass takes into account the amount of curvature inside the star (through $\Xi_{\star}$), its integrated density and pressure, and depends explicitly on the range of the extra scalar degree of freedom. The dif\mbox{}ference between the two masses can be expressed as follows:
\begin{equation}
	M_g - M_{\star} = \bigg( \frac{\tilde{M}_{\star}}{3} - \tilde{P}_{\star} \bigg) \, e^{-mr_\star} \big( 1 + m r_\star \big) - \frac{M_{\star}}{3} + P_{\star} + \Xi_{\star} \; .
\end{equation}
It is not immediately evident which mass is larger, since both $\tilde{M}_{\star}/3 - \tilde{P}_{\star}$ and $M_{\star}/3 - P_{\star}$ are non-negative (for non-exotic matter at least).

In the $m \to \infty$ limit, which corresponds to the GR limit $\alpha\to 0\,$, from eq.~\eqref{chieq} we recover the GR result $\z/\a \to 8 \pi G_{_{\! N}} \big(\rho - 3p \big)$ and thus:
\begin{equation}
	\Xi_{\star} \xrightarrow[m \to \infty]{} \frac{M_{\star}}{3} - P_{\star} \qquad \mbox{and} \qquad M_g \xrightarrow[m \to \infty]{} M_{\star} \; ,
\end{equation} 
as expected.\footnote{The result $\z/\a \to 8 \pi G_{_{\! N}} \big(\rho - 3p \big)$ when $m \to \infty$ can also be obtained from the expression (\ref{alphachisolutionfull}), by recognizing an appropriate realization of the Dirac delta inside the integral.} In the opposite limit $m \to 0\,$, which means that the range of $\z$ tends to inf\mbox{}inity, taking into account (\ref{MCandPdefinitionsmrsmall}) we get
\begin{equation}
	M_g - M_{\star} \xrightarrow[m \to 0]{} \Xi_{\star} \; .
\end{equation}
Since the expression (\ref{alphachiapproxfull}) implies that $\z$ remains bounded when $m \to 0\,$, from the def\mbox{}inition (\ref{Xidefinition}) it follows that in this limit $\Xi_{\star} \to 0$ and therefore
\begin{equation} \label{Mass difference m small}
	M_g - M_{\star} \xrightarrow[m \to 0]{} 0 \; .
\end{equation}
In the forthcoming paper \cite{forthcoming} we discuss the magnitude and physical meaning of the dif\mbox{}ference between $M_g$ and $M_{\star}$, focusing for the sake of concreteness on neutron stars and their mass-radius relation.

\subsubsection{The PPN gamma parameter}

The second interesting conclusion that we draw from our analysis is a prediction on the post-Newtonian parameter $\gamma$
\beq \label{gamma def}
	\g = \frac{\Psi}{\Phi} \; ,
\eeq
which from eqs.~\eqref{Psisolexternal1} and \eqref{Phisolutionexternal1} can be expressed as follows:
\begin{equation}
		\gamma = \frac{3 \, r_{\!g} - r_{\!\z} \, e^{-mr} (1 + mr)}{3 \, r_{\!g} + r_{\!\z} \, e^{-mr}} \; ,
\end{equation}
and explicitly
\begin{equation} \label{gammalin}
		\gamma = \frac{\big( \tilde{M}_{\star} - 3 \tilde{P}_{\star} \big) \, e^{-m r_{\star}} (1 + m r_{\star}) + 2 M_{\star} + 3 P_{\star} + 3 \, \Xi_{\star} - \big( \tilde{M}_{\star} - 3 \tilde{P}_{\star} \big) \, e^{-mr} (1 + mr)}{\big( \tilde{M}_{\star} - 3 \tilde{P}_{\star} \big) \, e^{-m r_{\star}} (1 + m r_{\star}) + 2 M_{\star} + 3 P_{\star} + 3 \, \Xi_{\star} + \big( \tilde{M}_{\star} - 3 \tilde{P}_{\star} \big) \, e^{-mr}} \; .
\end{equation}
Well outside the range of the scalar degree of freedom, that is for radii $r \gg m^{-1}$, we recover $\gamma = 1$ as expected (the spacetime being asymptotically f\mbox{}lat).

Let us however suppose that $m r_{\star} \ll 1$, and consider the region well inside the range of the scalar degree of freedom (that is, $r_{\star} \leq r \ll m^{-1}$). Recalling eq.~\eqref{MCandPdefinitionsmrsmall} and the fact that $\Xi \to 0$ when $m \to 0\,$, we get:
\begin{equation} \label{gamma nearby}
	\gamma \simeq \frac{2 M_{\star} + 3 P_{\star}}{4 M_{\star} - 3 P_{\star}} \qquad \mbox{for} \quad r_{\star} \leq r \ll m^{-1} \; ,
\end{equation}
which in general is smaller than $1$ but larger than $1/2$, unless the pressure vanishes. This relation generalizes the results of various papers in the literature, for example \cite{Chiba:2006jp, Olmo:2006eh, Kainulainen:2007bt}, which f\mbox{}ind $\g = 1/2$ by neglecting the pressure. It is worthwhile to recall that the condition of asymptotic f\mbox{}latness is consistent for models where $f$ is analytic in $R = 0\,$, while this is not true in models like, for example, $f(R) = R - \m^{4}/R\,$. Therefore, our result should be properly seen as a generalization of the results concerning the former class of $f(R)$ models.

\subsubsection{Relativistic (conformal) matter}
\label{Relativistic matter}

Note that, for a hypothetical star composed exclusively of relativistic matter and radiation, we obtain $\g = 1$ independently from the range of the extra scalar degree of freedom. This agrees with our previous comment at the end of section \ref{External and internal solutions}, about the extra degree of freedom being masked. Interestingly, we may expect the central part of a neutron star to be described by the equation of state $p = \r/3$ when the central density is high enough, although with certainty it won't be so for the outer layers \cite{Weinberg:1972}. While this means that the masking of the scalar d.o.f.\ cannot be relevant for ordinary stars, it leaves open the possibility of this behaviour being relevant for exotic stars. In particular, the Bag Model for quark stars \cite{Chodos:1974je, Peshier:1999ww, Alford:2004pf, Schmitt:2010pn} predicts an equation of state of the form $p = \rho/3 + B\,$, where $B$ is the Bag Model constant, whose value is taken to be about $B \sim 100 \; \text{MeV}^4$. In that case
\begin{equation}
\gamma = \frac { M_{\star} + B_{\star}} { M_{\star} - B_{\star}} \quad ,
\end{equation}
where  $B_{\star} \equiv \int_{V_{\star}} B \, dV \,$. Considering a star with the same mass of the sun $ M_{\star} \sim M_{_{\odot}}$, we get
\begin{equation}
\gamma \sim 1.2 \quad .
\end{equation}

Both the conclusions $\gamma = 1$ and $\z = 0$ could have been foretold directly from the fourth order equation (\ref{metric eq}). Let us consider the case of conformal matter (i.e.\ matter whose energy-momentum tensor is trace-free). Taking f\mbox{}irst of all the trace of the equation (\ref{metric eq}), we get the following second order equation for $R$:
\begin{equation}\label{traced metric eq conformal}
	\Box \, R - m^{2} R = 0 \; .
\end{equation}
Considering static conf\mbox{}igurations, the only regular and asymptotically decaying solution of the latter equation is $R = 0\,$. This conclusion relies crucially on the condition $T = 0$ being valid everywhere (i.e.\ also at the origin). Inserting $R = 0$ back into the equation (\ref{metric eq}) we get:
\begin{equation}\label{metric eq conformal}
	R_{\m\n} = 8 \pi G_{_{\! N}} \, T_{\m\n} \; ,
\end{equation}
which is Einsten's equation in presence of conformal matter. This shows that, at least for static, regular and asymptotically f\mbox{}lat conf\mbox{}igurations, Starobinsky $f(R)$ gravity gives the same predictions as GR when only conformal matter is present. 

\section{Numerical solutions}
\label{Numerical solutions}

In this section we support our analytic study in the linear approximation with fully numerical solutions of the non-linear system of equations \eqref{systemTOV}. The part which especially calls for conf\mbox{}irmation is section \ref{Consistency internal solutions}, which concerns our assumptions about the behaviour of the f\mbox{}ields $\Phi$, $\Psi$ and $\z$ inside the star. While we showed that, outside the star, our approximations follow from the weak-f\mbox{}ield conditions $\abs{\Phi} \ll 1\,$, $\abs{\Psi} \ll 1\,$, and $\abs{\z} \ll 1\,$, regarding the interior of the star we just assumed this to be the case. It is important to understand whether this indeed happens, and whether it is a quite general property or not. To investigate this, and in particular probe the generality of our assumptions, it is convenient to consider a star conf\mbox{}iguration which is close to the limit of validity of the linear approximation. For this reason we decide to use neutron stars as a testbench for our analysis.

This choice may seem questionable, since neutron stars are naively associated with the idea of the gravitational f\mbox{}ield being ``strong''. Recall however that, for what concerns static and spherically symmetric solutions in GR, the non-linear terms in the equations of motion become of the same order of the linear ones at the Schwarzschild radius, which means that outside a (static and spherically symmetric) neutron star the former are subdominant (although in general not negligible). To compromise 
with the request of the linear approximation being valid, we consider neutron stars with suitably small central density. We model our star with the polytropic equation of state:
\begin{equation}
	p = k \rho^{2} \; ,
\end{equation}
with
\begin{equation}
	k = 4.012 \times 10^{-4} \; \mbox{fm}^{3}/\mbox{MeV} \; .
\end{equation}
The choice of this equation of state, which admittedly provides only an approximated description of a neutron star \cite{Silbar:2003wm}, is justif\mbox{}ied because our main aim here is testing the validity of the analysis of the sections \ref{Linearization} and \ref{Linear gravity}. For this reason we don't feel necessary to use a more realistic Sly equation of state \cite{Douchin:2001sv}.

\subsection{Numerical strategy}

As a preparation to the numerical integration, we recast the system~\eqref{systemTOV} in terms of dimensionless quantities. To this aim, we choose to normalize the quantities under consideration to the Sun's half Schwarzschild radius:
\begin{equation}
	r_{_{\!0}} \equiv \frac{G_{_{\! N}} M_{_{\odot}}}{c^2} \simeq 1.5 \; \mbox{km} \; ,
\end{equation}
and to provide a transparent dimensional analysis we temporarily reinstate the constant $c\,$. We therefore introduce the dimensionless radial coordinate $x = r/r_{_{\!0}}\,$, and work with the following dimensionless quantities:
\begin{align}
	\hat{\r}(x)&= \frac{G_{_{\! N}} r_{_{\!0}}^2}{c^4} \, \rho \big( r_{_{\!0}} x \big) \; , & \hat{p}(x)&= \frac{G_{_{\! N}} r_{_{\!0}}^2}{c^4} \, p \big( r_{_{\!0}} x \big) \; , & \hat{\a} =&\frac{\alpha}{r_{_{\!0}}^2} \; , & \hat{k} =&\frac{c^{4}}{G_{_{\! N}} r_{_{\!0}}^2} \, k \simeq 139 \; ,
\end{align}
along with $\hat{\z}(x) = \z(r_{_{\!0}} x)\,$, $\hat{A}(x) = A(r_{_{\!0}} x)$ and $\hat{B}(x) = B(r_{_{\!0}} x)\,$.

To f\mbox{}ind the numerical solutions we employ a shooting method, imposing the initial conditions at the center of the star. Note on this respect that, although the equations \eqref{systemTOV} are formally singular at $r = 0\,$, they admit solutions which are regular at the origin. It is not dif\mbox{}f\mbox{}icult to see that, asking the pressure and the density at the center of the star to be f\mbox{}inite, solutions which are analytic in $r = 0$ can exist only provided
\begin{align}
	A^{\p}(0) &= 0 & B(0) &= 1 & B^{\p}(0) &= 0 & \z^{\p}(0) &= 0 \quad .
\end{align}
Furthermore, the equations \eqref{systemTOV} are invariant under the rescaling of $A$ by an arbitrary constant. It follows that changing the initial condition $A(0) = A_{_{0}}$ results simply in a rescaling of the function $A(r)$, while leaving $B(r)$ and $\z(r)$ unchanged. We are therefore free to choose the initial condition $A_{_{0}} = 1$, and a posteriori rescale the function $A(r)$ such that it tends to 1 far from the star. This leads us to consider the conditions
\begin{align} \label{initial conditions origin}
	\hat{A}(0) &= 1 & \hat{B}(0) &= 1 & \hat{\z}(0) &= \hat{\z}_{_{0}} & \frac{d \hat{\z}}{dx}(0) &= 0 & \hat{\r}(0) &= \hat{\r}_{_{0}} \quad ,
\end{align}
where the pressure $\hat{p}$ has not been included since it is determined by $\hat{\r}$ via the equation of state. The central density $\hat{\r}_{_{0}}$ and central curvature $\hat{\z}_{_{0}}$ are free parameters. We choose the former a priori and then determine $\hat{\z}_{_{0}}$ via the shooting, selecting the solution for $\hat{\z}$ which decays exponentially to zero far from the star (meaning at $x \gg \sqrt{\hat{\a}} \sim (m r_{_{\!0}})^{-1}$) where the external solution eq.~\eqref{alphachisolutionmatter} is valid. The star's radius is determined to be that for which $\hat{\r}$ becomes negative.

There is a technical subtlety regarding the initial conditions. Since the formal singularity at $r = 0$ is problematic for the numerical integrator, it is not possible to set the initial conditions for the numerical code exactly at $x = 0\,$. We therefore use $x_i = 10^{-5}$ as the initial radius for the numerical code. Plugging into the equations \eqref{systemTOV} the Taylor expansion of $\hat{A}$, $\hat{B}$, $\hat{\z}$, $\hat{\r}$ and $\hat{p}$ around $x = 0\,$, and working at f\mbox{}irst order in $x_i\,$, it can be seen that the conditions (\ref{initial conditions origin}) imply the following initial conditions:
\begin{align}
	\hat{A}(x_i) &= 1 \; , & \hat{B}(x_i) &= 1 \; , & \hat{\r}(x_i) &= \hat{\r}_{_{0}} \; , & \hat{\z}(x_i) &= \hat{\z}_{_{0}} \; ,
\end{align}
and
\beq
	\frac{d \hat{\z}}{dx}(x_i) = \frac{1}{ 6 \times 10^5} \, \bigg[ \, \frac{\hat{\z}_{_{0}}}{\hat{\a}} + 8 \pi \hat{\r}_{_{0}} \Big( 3 \, \hat{k} \hat{\r}_{_{0}} - 1 \Big) \bigg] \; .
\eeq

\subsection{Numerical results}

We choose for def\mbox{}initeness the model with $\a = 10^7 \, r_{_{\!0}}^{2}$ and consider a neutron star with central density $\r_{_{0}} = 4 \times 10^{13} \, \mathrm{g/cm^{3}}$. The star's radius inferred from the shooting is $r_{\star} \simeq 11.60 \, r_{_{\!0}}$, while the range of the scalar degree of freedom is $m^{-1} \simeq 7.746 \times 10^{3} \, r_{_{\!0}}$, so the ratio of the former to the latter is $m r_{\star} \simeq 1.498 \times 10^{-3}$. This implies that we are in the regime where (\ref{gamma nearby}) should hold, that is the regime where the range of the scalar degree of freedom is much bigger than the star's radius.

\begin{figure}[hbtp]
 \subfigure[The gravitational potentials times $r/r_{_{0}}$ \label{fig:rPhirPsi}]{\includegraphics[width=0.645\columnwidth]{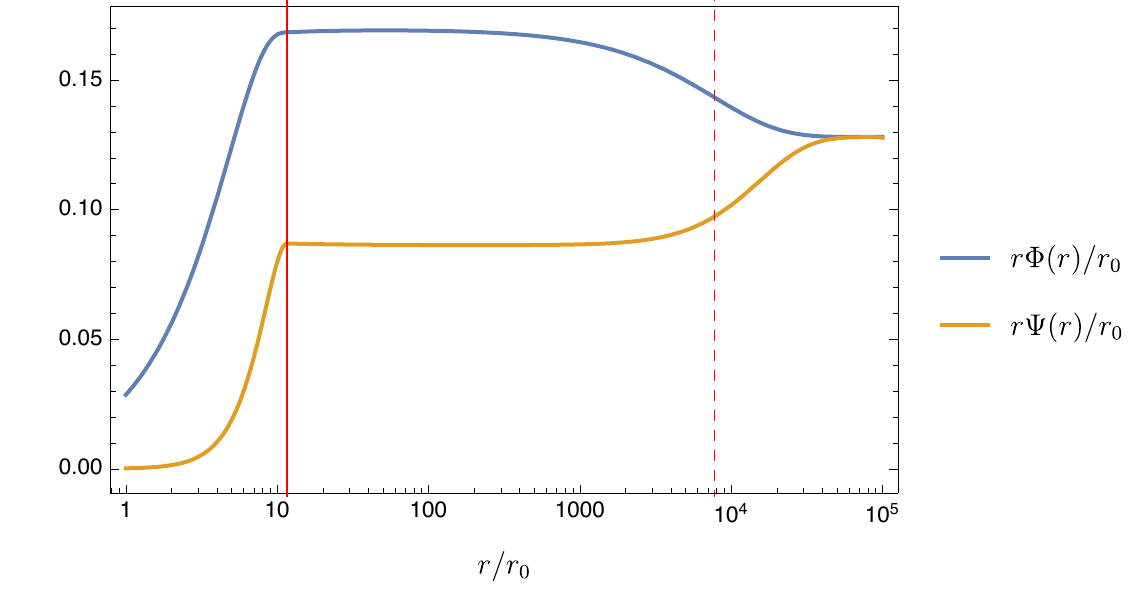}} 
 \subfigure[The PPN parameter $\g$ \label{fig:gammaPPN}]{\includegraphics[width=0.495\columnwidth]{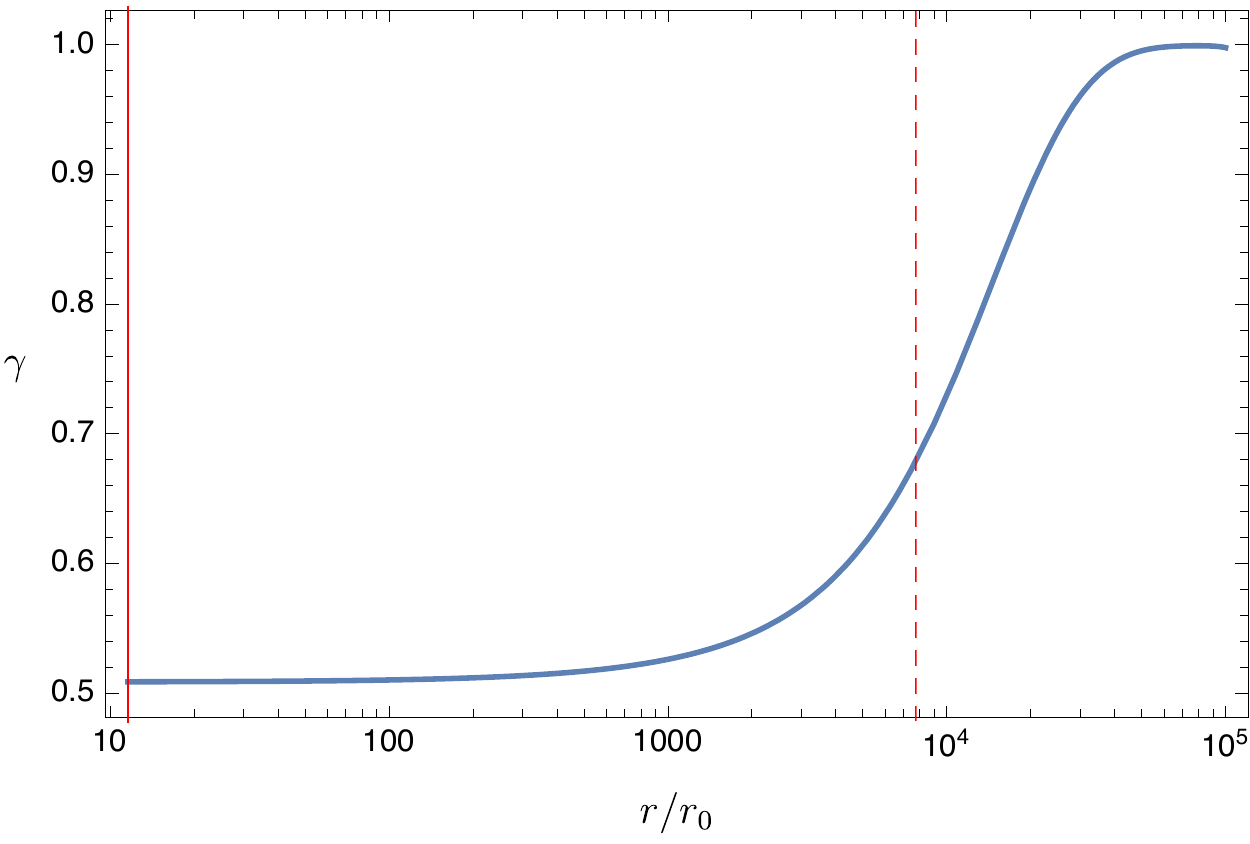}}
 \caption{The behaviour of the gravitational f\mbox{}ield. The continuous vertical line marks the star's surface while the dashed line marks the range of the scalar d.o.f.}
\end{figure}

The behaviour of the gravitational f\mbox{}ield is clearly illustrated in the f\mbox{}igures \ref{fig:rPhirPsi} and \ref{fig:gammaPPN}. In f\mbox{}igure \ref{fig:rPhirPsi} the gravitational potentials, multiplied by $r/r_{_{0}}\,$, are plotted. In agreement with the analysis of the previous sections, it is apparent that there are two Newtonian regions, one for $r_{\star} \leq r \ll m^{-1}$ and one for $r \gg m^{-1}$, with a non-Newtonian transition region in between. This behaviour is conf\mbox{}irmed by the plot of the PPN parameter $\g$, displayed in f\mbox{}igure \ref{fig:gammaPPN}, which tends asymptotically to unity as expected. Noteworthy, the value of the $\g$ parameter outside the star is $\g_{\star} \simeq 0.5085$, which is higher than $1/2$ at the 1\% level.

\begin{figure}[hbtp]
 \subfigure[The gravitational potentials \label{fig:PhiPsi}]{\includegraphics[width=0.6\columnwidth]{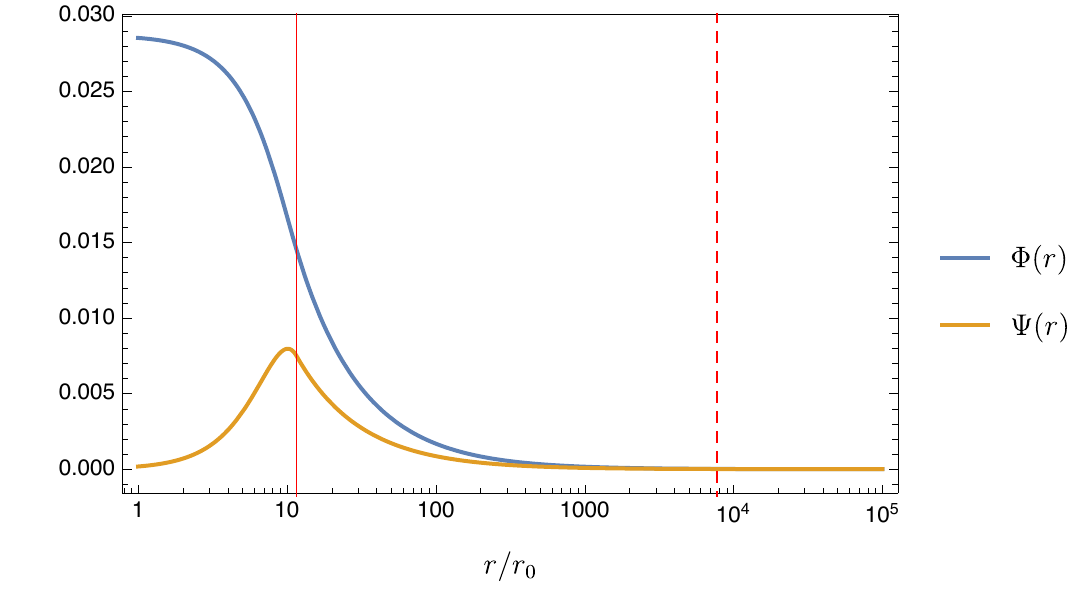}} \quad
 \subfigure[The scalar d.o.f.\ $\z$ \label{fig:zeta}]{\includegraphics[width=0.49\columnwidth]{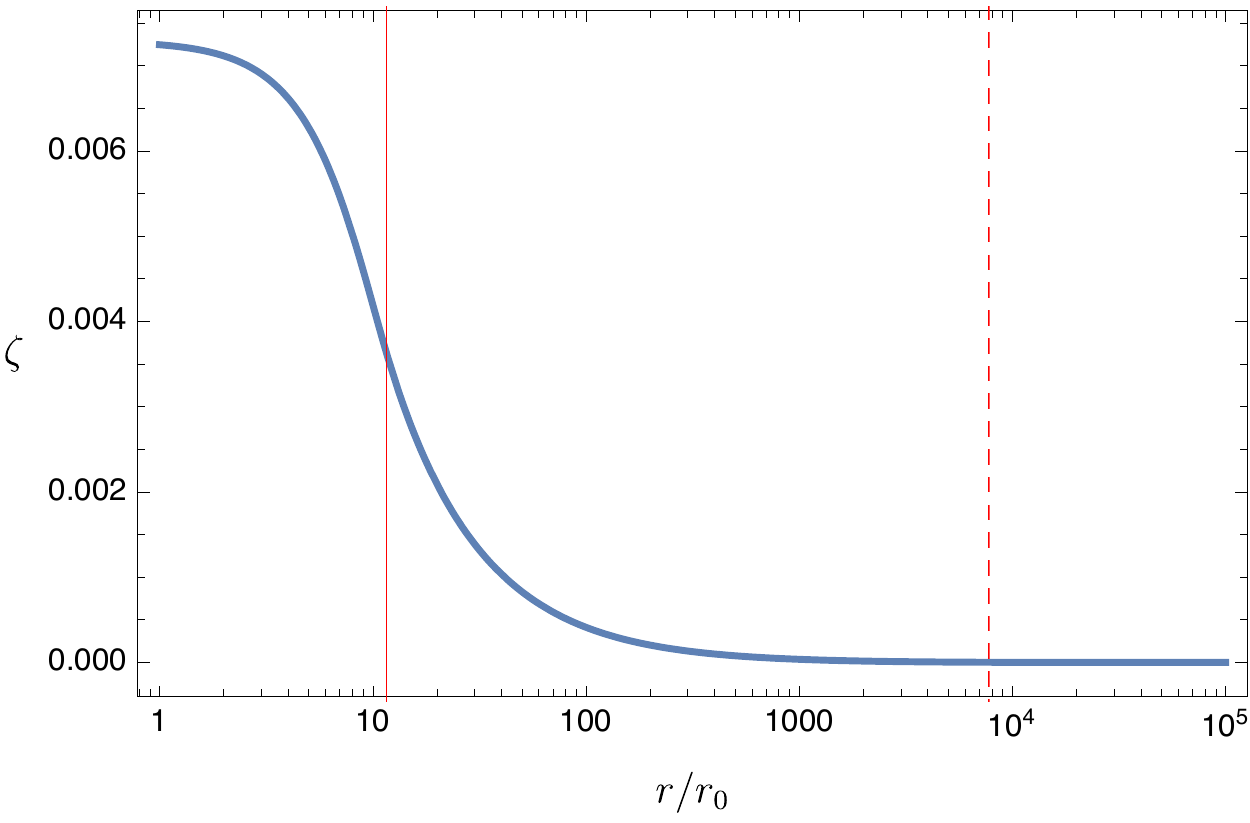}}
 \caption{The amplitude of the gravitational potentials and the scalar d.o.f.\ $\z$. The meaning of the vertical lines is the same as above.}
\end{figure}

In f\mbox{}igures \ref{fig:PhiPsi} and \ref{fig:zeta} the gravitational potentials (this time not multiplied by $r/r_{_{0}}$) and the scalar degree of freedom $\z$ are plotted. It is easy to see that they are everywhere of the order $10^{-2}$ or smaller, so they satisfy the weak f\mbox{}ield conditions $\abs{\Phi} \ll 1\,$, $\abs{\Psi} \ll 1$ and $\abs{\z} \ll 1\,$. In relation to the discussion of section \ref{Consistency approximation}, the numerically inferred gravitational radii are $r_{\!\z} = 0.2503 \, r_{_{\!0}}$ and $r_{\!g} = 0.2561 \, r_{_{\!0}}$, a factor $10^{2}$ smaller than the star's radius. According to our analysis, this should imply the conditions (\ref{second group}) to be satisf\mbox{}ied outside the star. Indeed, we verif\mbox{}ied numerically that all the terms appearing in (\ref{second group}) are of order $10^{-2}$ or smaller. Most importantly, this holds also \emph{inside} the star, which conf\mbox{}irms our assumption about the internal behaviour of the solutions (see section \ref{Consistency internal solutions}). In particular $\z$ is indeed monotonically decreasing, as suggested and motivated in the aforementioned section. Overall, these numerical results strongly support the consistency of our approximations and the validity of the analysis of sections \ref{Linearization} and \ref{Linear gravity}.

Let us comment on the relation (\ref{gamma nearby}). The numerical analysis gives the following estimates
\begin{align}
M_{\star} &= 0.128682 \, M_{_{\odot}} & P_{\star} &= 4.7918 \times 10^{-4} \, M_{_{\odot}} & \Xi_{\star} &= 4.11868 \times 10^{-8} \, M_{_{\odot}}
\end{align}
and indicates that $\tilde{M}_{\star}$ is equal to $M_{\star}$ to the f\mbox{}ifth signif\mbox{}icant digit (the same happens for $\tilde{P}_{\star}$ and $P_{\star}$). The smallness of $\Xi_{\star}$, $\tilde{M}_{\star} - M_{\star}$ and $\tilde{P}_{\star} - P_{\star}$ is expected from (\ref{MCandPdefinitionsmrsmall}) and from the analysis of section \ref{Relationship masses}, and implies that approximating (\ref{gammalin}) with (\ref{gamma nearby}) is justif\mbox{}ied in the context of the linear approximation.

\section{Conclusions and comments}
\label{Conclusions}

In this paper we studied static, spherically symmetric and asymptotically f\mbox{}lat conf\mbox{}igurations of Starobinsky $f(R)$ gravity in the metric formalism. We focused on the behaviour of gravity in the linear regime in presence of a static star, without a priori assuming the pressure of the latter to be negligible.

We found that the PPN parameter $\g$ outside the star and well inside the range of the scalar degree of freedom can be larger than $1/2\,$. We provided a relation which explicitly expresses $\g$ in terms of the mass, of the integrated pressure of the star and of the ratio between the star's radius and the range of the extra degree of freedom. We showed explicitly that at least two dif\mbox{}ferent, sensible notions of mass (frequently used in the literature for analysing e.g.\ the stability of neutron stars) are possible in the context of the $f(R)$ model under consideration (and possibly also for more general models). We derived an analytic formula relating the two, showing, as expected, that they coincide in the GR limit. Though aware that this might be an unrealistic conf\mbox{}iguration, we also showed that in the limit where the matter content forming the star tends to pure radiation, then no ef\mbox{}fect from the extra scalar degree of freedom appears, as if it were masked.

Our analysis and results are based on a set of assumptions, which include the usual weak-f\mbox{}ield conditions, whose consistency was analytically proved only for the exterior solutions of the f\mbox{}ields. The interior ones are described by semi-analytic formulae and depend on integrals which can be computed, in practice, only numerically. Therefore, in order to assess the viability of our assumptions regarding the interior of the star, we performed a numerical analysis considering for concreteness the case of a neutron star. We found that the conditions we assumed to hold inside the star are indeed satisf\mbox{}ied, thus providing f\mbox{}irm ground to our analytic study and results also in that range of radii.

It is wortwhile at this point to discuss how our analysis, and in particular our choice of considering asymptotically f\mbox{}lat conf\mbox{}igurations, relates to that of \cite{Chiba:2006jp}, where a (homogeneous) cosmological background, with associated cosmological energy-momentum tensor, is included. Note that in \cite{Chiba:2006jp}, whenever the gravitational f\mbox{}ield outside a star is studied, the time dependence of the cosmological background is neglected (with the sensible justif\mbox{}ication that the cosmological evolution is slow compared to the typical time-scales of, say, the solar system). In this case, therefore, the role of the cosmological background reduces purely to setting the boundary conditions for the metric and the scalar curvature, when the distance from the star tends to inf\mbox{}inity. Allowing for a non-zero cosmological value $R_0$ for the scalar curvature is indeed crucial for a generic $f(R)$ theory, since $R = 0$ is not necessarily an allowed conf\mbox{}iguration of the theory (for example, it is not allowed if $f(R)$ contains negative powers of $R$). However, for the Starobinsky model the cosmological background given by $R_0 = 0$ \emph{is} an allowed conf\mbox{}iguration. As a matter of fact, when the Starobinsky model is studied in \cite{Chiba:2006jp}, they set to zero both the cosmological energy-momentum tensor and the background value of the scalar curvature, that is they take $R_0 = 0\,$. This shows that considering asymptotically f\mbox{}lat solutions not only is compatible with the approach of \cite{Chiba:2006jp}, being equivalent to considering a homogeneous and static cosmological background with $R_0 = 0$, but actually f\mbox{}its nicely in that approach.

A natural further development of this work would be to generalise its results for a generic function $f(R)$, at least qualitatively (since a numerical analysis without specifying a functional form for $f(R)$ is impossible). Moreover, in a forthcoming paper \cite{forthcoming} we are analysing a realistic model of neutron star, focusing on the mass-radius relation, in order to understand how the use of the two def\mbox{}initions of mass introduced in the present paper inf\mbox{}luences the analysis of the properties of a neutron star and their estimation from observational data.

\section*{Acknowledgments}
FS acknowledges partial f\mbox{}inancial support from CNPq (Brazil) and FAPES (Brazil). OFP thanks the Alexander von Humboldt foundation for funding and the Institute for Theoretical Physics of the Heidelberg University for kind hospitality. OFP also acknowledges partial f\mbox{}inancial support from CNPq, FAPES and CAPES. This study was f\mbox{}inanced in part by the \emph{Coordena\c{c}\~ao de Aperfei\c{c}oamento de Pessoal de N\'ivel Superior} - Brazil (CAPES) - Finance Code 001. SEJ ackowledges f\mbox{}inancial support from UFES in the occasion of a visit to the Astrophysics, Cosmology and Gravitation group.


\begin{thebibliography}{99}

\bibitem{Sotiriou:2008rp}
T.~P.~Sotiriou and V.~Faraoni,
\newblock {\em f(R) Theories Of Gravity},
\newblock {\em Rev.\ Mod.\ Phys.\ }\textbf{82} (2010) 451--497, doi:10.1103/RevModPhys.82.451, \texttt{0805.1726[gr-qc]}.

\bibitem{DeFelice:2010aj}
A.~De~Felice and S.~Tsujikawa,
\newblock {\em f(R) theories},
\newblock {\em Living Rev.\ Rel.\ }\textbf{13} (2010) 3, doi:10.12942/lrr-2010-3, \texttt{1002.4928[gr-qc]}.

\bibitem{Nojiri:2010wj}
S.~Nojiri and S.D.~Odintsov,
\newblock {\em Unif\mbox{}ied cosmic history in modif\mbox{}ied gravity: from F(R) theory to Lorentz non-invariant models},
\newblock {\em Phys.\ Rept.\ }\textbf{505} (2011) 59--144, doi:10.1016/j.physrep.2011.04.001, \texttt{1011.0544[gr-qc]}.

\bibitem{Nojiri:2017ncd}
S.~Nojiri, S.D.~Odintsov and V.K.~Oikonomou,
\newblock {\em Modif\mbox{}ied Gravity Theories on a Nutshell: Inf\mbox{}lation, Bounce and Late-time Evolution},
\newblock {\em Phys.\ Rept.\ }\textbf{692} (2017) 1--104, doi:10.1016/j.physrep.2017.06.001, \texttt{1705.11098[gr-qc]}.

\bibitem{Buchdahl1970}
H.~A.~Buchdahl,
\newblock {\em Non-linear Lagrangians and cosmological theory},
\newblock {\em Mon.\ Not.\ Roy.\ Astron.\ Soc.\ }\textbf{150} (1970) 1.

\bibitem{Ostrogradsky:1850fid}
M.~Ostrogradsky,
\newblock {\em M\'emoires sur les \'equations dif\mbox{}f\'erentielles, relatives au probl\`eme des isop\'erim\`etres},
\newblock {\em Mem.\ Acad.\ St.\ Petersbourg} \textbf{6} (1850) no.4, 385--517.

\bibitem{Woodard:2006nt}
R.~P.~Woodard,
\newblock {\em Avoiding dark energy with 1/r modif\mbox{}ications of gravity},
\newblock {\em Lect.\ Notes Phys.\ }\textbf{720} (2007) 403--433, doi:10.1007/978-3-540-71013-4\_14, \texttt{[astro-ph/0601672]}.

\bibitem{Dolgov:2003px}
A.~D.~Dolgov and M.~Kawasaki,
\newblock {\em Can modif\mbox{}ied gravity explain accelerated cosmic expansion?},
\newblock {\em Phys.\ Lett.\ }\textbf{B 573} (2003) 1--4, doi:10.1016/j.physletb.2003.08.039, \texttt{[astro-ph/0307285]}.

\bibitem{Khoury:2003aq}
J.~Khoury and A.~Weltman,
\newblock {\em Chameleon f\mbox{}ields: Awaiting surprises for tests of gravity in space},
\newblock {\em Phys.\ Rev.\ Lett.\ }\textbf{93} (2004) 171104, doi:10.1103/PhysRevLett.93.171104, \texttt{[astro-ph/0309300]}.

\bibitem{Khoury:2003rn}
J.~Khoury and A.~Weltman,
\newblock {\em Chameleon cosmology},
\newblock {\em Phys.\ Rev.\ }\textbf{D 69} (2004) 044026, doi:10.1103/PhysRevD.69.044026, \texttt{[astro-ph/0309411]}.

\bibitem{Accioly:1998nm}
A.~Accioly, A.~D.~Azeredo, E.~C.~de~Rey~Neto and H.~Mukai,
\newblock {\em Bending of light in the framework of R + R**2 gravity},
\newblock {\em Braz.\ J.\ Phys.\ }\textbf{28} (1998) 496--504.

\bibitem{Accioly:2000gd}
A.~Accioly, H.~Blas and H.~Mukai,
\newblock {\em Light def\mbox{}lection and quadratic gravity},
\newblock {\em Nuovo Cim.\ }\textbf{B 115} (2000) 1235--1239.

\bibitem{Dick:2003dw}
R.~Dick,
\newblock {\em On the Newtonian limit in gravity models with inverse powers of R},
\newblock {\em Gen.\ Rel.\ Grav.\ }\textbf{36} (2004) 217--224, doi:10.1023/B:GERG.0000006968.53367.59, \texttt{[gr-qc/0307052]}.

\bibitem{Nojiri:2003ft}
S.~Nojiri and S.D.~Odintsov,
\newblock {\em Modif\mbox{}ied gravity with negative and positive powers of the curvature: Unif\mbox{}ication of the inf\mbox{}lation and of the cosmic acceleration},
\newblock {\em Phys.\ Rev.\ }\textbf{D 68} (2003) 123512, doi:10.1103/PhysRevD.68.123512, \texttt{[hep-th/0307288]}.

\bibitem{Chiba:2003ir}
T.~Chiba,
\newblock {\em 1/R gravity and scalar-tensor gravity},
\newblock {\em Phys.\ Lett.\ }\textbf{B 575} (2003) 1-3, doi:10.1016/j.physletb.2003.09.033, \texttt{[astro-ph/0307338]}.

\bibitem{Soussa:2003re}
M.~E.~Soussa and R.~P.~Woodard,
\newblock {\em The force of gravity from a Lagrangian containing inverse powers of the Ricci scalar},
\newblock {\em Gen.\ Rel.\ Grav.\ }\textbf{36} (2004) 855--862, doi:10.1023/B:GERG.0000017037.92729.69, \texttt{[astro-ph/0308114]}.

\bibitem{Rajaraman:2003st}
A.~Rajaraman,
\newblock {\em Newtonian gravity in theories with inverse powers of R}
\newblock (2003), \texttt{[astro-ph/0311160]}.

\bibitem{Easson:2004fq}
D.~A.~Easson,
\newblock {\em Cosmic acceleration and modif\mbox{}ied gravitational models},
\newblock {\em Int.\ J.\ Mod.\ Phys.\ }\textbf{A 19} (2004) 5343--5350, doi:10.1142/S0217751X04022578, \texttt{[astro-ph/0411209]}.

\bibitem{Olmo:2005zr}
G.~J.~Olmo,
\newblock {\em The Gravity Lagrangian according to solar system experiments},
\newblock {\em Phys.\ Rev.\ Lett.\ }\textbf{95} (2005) 261102, doi:10.1103/PhysRevLett.95.261102, \texttt{[gr-qc/0505101]}.

\bibitem{Olmo:2005hc}
G.~J.~Olmo,
\newblock {\em Post-Newtonian constraints on f(R) cosmologies in metric and Palatini formalism},
\newblock {\em Phys.\ Rev.\ }\textbf{D 72} (2005) 083505, doi:10.1103/PhysRevD.72.083505, \texttt{[gr-qc/0505135]}.

\bibitem{Navarro:2005gh}
I.~Navarro and K.~Van~Acoleyen,
\newblock {\em On the Newtonian limit of generalized modif\mbox{}ied gravity models},
\newblock {\em Phys.\ Lett.\ }\textbf{B 622} (2005) 1--5, doi:10.1016/j.physletb.2005.07.008, \texttt{[gr-qc/0506096]}.

\bibitem{Cembranos:2005fi}
J.~A.~R.~Cembranos,
\newblock {\em The Newtonian limit at intermediate energies},
\newblock {\em Phys.\ Rev.\ }\textbf{D 73} (2006) 064029, doi:10.1103/PhysRevD.73.064029, \texttt{[gr-qc/0507039]}.

\bibitem{Capozziello:2005bu}
S.~Capozziello and A.~Troisi,
\newblock {\em PPN-limit of fourth order gravity inspired by scalar-tensor gravity},
\newblock {\em Phys.\ Rev.\ }\textbf{D 72} (2005) 044022, doi:10.1103/PhysRevD.72.044022, \texttt{[astro-ph/0507545]}.

\bibitem{Clifton:2005aj}
T.~Clifton and J.~D.~Barrow,
\newblock {\em The Power of General Relativity},
\newblock {\em Phys.\ Rev.\ }\textbf{D 72} (2005) no.10 103005, doi:10.1103/PhysRevD.72.103005
\newblock [\emph{Erratum ibid} \textbf{D 90} (2014) no.2 029902, doi:10.1103/PhysRevD.90.029902], \texttt{[gr-qc/0509059]}.

\bibitem{Barrow:2005dn}
J.~D.~Barrow and T.~Clifton,
\newblock {\em Exact cosmological solutions of scale-invariant gravity theories},
\newblock {\em Class.\ Quant.\ Grav.\ }\textbf{23} (2006) L1, doi:10.1088/0264-9381/23/1/L01, \texttt{[gr-qc/0509085]}.

\bibitem{Shao:2005wt}
C.-G.~Shao, R.-G.~Cai, B.~Wang and R.-K.~Su,
\newblock {\em An Alternative explanation of the conf\mbox{}lict between 1/R gravity and solar system tests},
\newblock {\em Phys.\ Lett.\ }\textbf{B 633} (2006) 164--166, doi:10.1016/j.physletb.2005.11.060, \texttt{[gr-qc/0511034]}.

\bibitem{Navarro:2005da}
I.~Navarro and K.~Van~Acoleyen,
\newblock {\em Consistent long distance modif\mbox{}ication of gravity from inverse powers of the curvature},
\newblock {\em JCAP} \textbf{03} (2006) 008, doi:10.1088/1475-7516/2006/03/008, \texttt{[gr-qc/0511045]}.

\bibitem{Capozziello:2006jj}
S.~Capozziello, A.~Stabile and A.~Troisi,
\newblock {\em Fourth-order gravity and experimental constraints on Eddington parameters},
\newblock {\em Mod.\ Phys.\ Lett.\ }\textbf{A 21} (2006) 2291--2301, doi:10.1142/S0217732306021633, \texttt{[gr-qc/0603071]}.

\bibitem{Multamaki:2006zb}
T.~Multamaki and I.~Vilja,
\newblock {\em Spherically symmetric solutions of modif\mbox{}ied f\mbox{}ield equations in f(R) theories of gravity},
\newblock {\em Phys.\ Rev.\ }\textbf{D 74} (2006) 064022, doi:10.1103/PhysRevD.74.064022, \texttt{[astro-ph/0606373]}.

\bibitem{Jin:2006if}
X.-H.~Jin, D.-J.~Liu and X.-Z.~Li,
\newblock {\em Solar System tests disfavor f(R) gravities}
\newblock (2006) \texttt{[astro-ph/0610854]}.

\bibitem{Ruggiero:2006qv}
M.~L.~Ruggiero and L.~Iorio,
\newblock {\em Solar System planetary orbital motions and f(R) theories of gravity},
\newblock {\em JCAP} \textbf{01} (2007) 010, doi:10.1088/1475-7516/2007/01/010, \texttt{[gr-qc/0607093]}.

\bibitem{Navarro:2006mw}
I.~Navarro and K.~Van~Acoleyen,
\newblock {\em f(R) actions, cosmic acceleration and local tests of gravity},
\newblock {\em JCAP} \textbf{02} (2007) 022, doi:10.1088/1475-7516/2007/02/022, \texttt{[gr-qc/0611127]}.

\bibitem{Baghram:2007df}
S.~Baghram, M.~Farhang and S.~Rahvar,
\newblock {\em Modif\mbox{}ied gravity with f(R) = square root of R**2 - R0**2},
\newblock {\em Phys.\ Rev.\ }\textbf{D 75} (2007) 044024, doi:10.1103/PhysRevD.75.044024, \texttt{[astro-ph/0701013]}.

\bibitem{Zhang:2007ne}
P.-J.~Zhang,
\newblock {\em The behavior of f(R) gravity in the solar system, galaxies and clusters},
\newblock {\em Phys.\ Rev.\ }\textbf{D 76} (2007) 024007, doi:10.1103/PhysRevD.76.024007, \texttt{[astro-ph/0701662]}.

\bibitem{Capozziello:2007wc}
S.~Capozziello, A.~Stabile and A.~Troisi,
\newblock {\em Spherically symmetric solutions in f(R)-gravity via Noether Symmetry Approach},
\newblock {\em Class.\ Quant.\ Grav.\ }\textbf{24} (2007) 2153--2166, doi:10.1088/0264-9381/24/8/013, \texttt{[gr-qc/0703067]}.

\bibitem{Hu:2007nk}
W.~Hu and I.~Sawicki,
\newblock {\em Models of f(R) Cosmic Acceleration that Evade Solar-System Tests},
\newblock {\em Phys.\ Rev.\ }\textbf{D 76} (2007) 064004, doi:10.1103/PhysRevD.76.064004, \texttt{0705.1158[astro-ph]}.

\bibitem{Henttunen:2007bz}
K.~Henttunen, T.~Multamaki and I.~Vilja,
\newblock {\em Stellar conf\mbox{}igurations in f(R) theories of gravity},
\newblock {\em Phys.\ Rev.\ }\textbf{D 77} (2008) 024040, doi:10.1103/PhysRevD.77.024040, \texttt{0705.2683[astro-ph]}.

\bibitem{Nojiri:2007as}
S.~Nojiri and S.D.~Odintsov,
\newblock {\em Unifying inf\mbox{}lation with LambdaCDM epoch in modif\mbox{}ied f(R) gravity consistent with Solar System tests},
\newblock {\em Phys.\ Lett.\ }\textbf{B 657} (2007) 238-245, doi:10.1016/j.physletb.2007.10.027, \texttt{0707.1941[hep-th]}.

\bibitem{Capozziello:2007ms}
S.~Capozziello, A.~Stabile and A.~Troisi,
\newblock {\em The Newtonian Limit of f(R) gravity},
\newblock {\em Phys.\ Rev.\ }\textbf{D 76} (2007) 104019, doi:10.1103/PhysRevD.76.104019, \texttt{0708.0723[gr-qc]}.

\bibitem{Capozziello:2007id}
S.~Capozziello, A.~Stabile and A.~Troisi,
\newblock {\em Spherical symmetry in f(R)-gravity},
\newblock {\em Class.\ Quant.\ Grav.\ }\textbf{25} (2008) 085004, doi:10.1088/0264-9381/25/8/085004, \texttt{0709.0891[gr-qc]}.

\bibitem{Multamaki:2007jk}
T.~Multamaki and I.~Vilja,
\newblock {\em Constraining Newtonian stellar conf\mbox{}igurations in f(R) theories of gravity},
\newblock {\em Phys.\ Lett.\ }\textbf{B 659} (2008) 843--846, doi:10.1016/j.physletb.2007.12.022, \texttt{0709.3422[astro-ph]}.

\bibitem{Capozziello:2007eu}
S.~Capozziello and S.~Tsujikawa,
\newblock {\em Solar system and equivalence principle constraints on f(R) gravity by chameleon approach},
\newblock {\em Phys.\ Rev.\ }\textbf{D 77} (2008) 107501, doi:10.1103/PhysRevD.77.107501, \texttt{0712.2268[gr-qc]}.

\bibitem{Cognola:2007zu}
G.~Cognola, E.~Elizalde, S.~Nojiri, S.D.~Odintsov, L.~Sebastiani and S.~Zerbini,
\newblock {\em A Class of viable modif\mbox{}ied f(R) gravities describing inf\mbox{}lation and the onset of accelerated expansion},
\newblock {\em Phys.\ Rev.\ }\textbf{D 77} (2008) 046009, doi:10.1103/PhysRevD.77.046009, \texttt{0712.4017[hep-th]}.

\bibitem{Capozziello:2009vr}
S.~Capozziello, A.~Stabile and A.~Troisi,
\newblock {\em A General solution in the Newtonian limit of f(R)-gravity},
\newblock {\em Mod.\ Phys.\ Lett.\ }\textbf{A 24} (2009) 659--665, doi:10.1142/S0217732309030382, \texttt{0901.0448[gr-qc]}.

\bibitem{Capozziello:2010iha}
S.~Capozziello, A.~Stabile and A.~Troisi,
\newblock {\em The Post-Minkowskian limit of f(R)-gravity},
\newblock {\em Int.\ J.\ Theor.\ Phys.\ }\textbf{49} (2010) 1251--1261, doi:10.1007/s10773-010-0307-4, \texttt{1001.0847[gr-qc]}.

\bibitem{Capozziello:2010wt}
S.~Capozziello, A.~Stabile and A.~Troisi,
\newblock {\em Comparing scalar-tensor gravity and f(R)-gravity in the Newtonian limit},
\newblock {\em Phys.\ Lett.\ }\textbf{B 686} (2010) 79--83, doi:10.1016/j.physletb.2010.02.042, \texttt{1002.1364[gr-qc]}.

\bibitem{Berry:2011pb}
C.~P.~L.~Berry and J.~R.~Gair,
\newblock {\em Linearized f(R) Gravity: Gravitational Radiation and Solar System Tests},
\newblock {\em Phys.\ Rev.\ }\textbf{D 83} (2011) 104022, doi:10.1103/PhysRevD.83.104022
\newblock [\emph{Erratum ibid} \textbf{D 85} (2012) 089906, doi:10.1103/PhysRevD.85.089906], \texttt{1104.0819[gr-qc]}.

\bibitem{Chiba:2006jp}
T.~Chiba, T.~L.~Smith and A.~L.~Erickcek,
\newblock {\em Solar System constraints to general f(R) gravity},
\newblock {\em Phys.\ Rev.\ }\textbf{D 75} (2007) 124014, doi:10.1103/PhysRevD.75.124014, \texttt{[astro-ph/0611867]}.

\bibitem{Olmo:2006eh}
G.~J.~Olmo,
\newblock {\em Limit to general relativity in f(R) theories of gravity},
\newblock {\em Phys.\ Rev.\ }\textbf{D 75} (2007) 023511, doi:10.1103/PhysRevD.75.023511, \texttt{[gr-qc/0612047]}.

\bibitem{Kainulainen:2007bt}
K.~Kainulainen, J.~Piilonen, V.~Reijonen and D.~Sunhede,
\newblock {\em Spherically symmetric spacetimes in f(R) gravity theories},
\newblock {\em Phys.\ Rev.\ }\textbf{D 76} (2007) 024020, doi:10.1103/PhysRevD.76.024020, \texttt{0704.2729[gr-qc]}.

\bibitem{Guo:2013fda}
J.-Q.~Guo,
\newblock {\em Solar system tests of f(R) gravity},
\newblock {\em Int.\ J.\ Mod.\ Phys.\ }\textbf{D 23} (2014) no.4, 1450036, doi:10.1142/S0218271814500369, \texttt{1306.1853[astro-ph]}.

\bibitem{Will:2005va}
C.~M.~Will,
\newblock {\em The Confrontation between General Relativity and experiment},
\newblock {\em Living Rev.\ Rel.\ }\textbf{9} (2006) 3, doi:10.12942/lrr-2006-3, \texttt{[gr-qc/0510072]}.

\bibitem{Bertolami:2013qaa}
O.~Bertolami, R.~March and J.~P\'aramos,
\newblock {\em Solar System constraints to nonminimally coupled gravity},
\newblock {\em Phys.\ Rev.\ }\textbf{D 88} (2013) 064019, doi:10.1103/PhysRevD.88.064019, \texttt{1306.1176[gr-qc]}.

\bibitem{Hohmann:2013rba}
M.~Hohmann, L.~Jarv, P.~Kuusk and E.~Randla,
\newblock {\em Post-Newtonian parameters $\gamma$ and $\beta$ of scalar-tensor gravity with a general potential},
\newblock {\em Phys.\ Rev.\ }\textbf{D 88} (2013) 084054, doi:10.1103/PhysRevD.88.084054
\newblock [\emph{Erratum ibid} \textbf{D 89} (2014) 069901], doi:10.1103/PhysRevD.89.069901, \texttt{1309.0031[gr-qc]}.

\bibitem{Hohmann:2017qje}
M.~Hohmann and A.~Sch\"arer,
\newblock {\em Post-Newtonian parameters $\gamma$ and $\beta$ of scalar-tensor gravity for a homogeneous gravitating sphere},
\newblock {\em Phys.\ Rev.\ }\textbf{D 96} (2017) 104026, doi:10.1103/PhysRevD.96.104026, \texttt{1708.07851 [gr-qc]}.

\bibitem{Starobinsky:1980te}
A.~A.~Starobinsky,
\newblock {\em A New Type of Isotropic Cosmological Models Without Singularity},
\newblock {\em Phys.\ Lett.\ }\textbf{B 91} (1980) 99--102, doi:10.1016/0370-2693(80)90670-X.

\bibitem{Martin:2013nzq}
J.~Martin, C.~Ringeval, R.~Trotta and V.~Vennin,
\newblock {\em The Best Inf\mbox{}lationary Models After Planck},
\newblock {\em JCAP} \textbf{03} (2014) 039, doi:10.1088/1475-7516/2014/03/039, \texttt{1312.3529[astro-ph]}.

\bibitem{Teyssandier:1983zz}
P.~Teyssandier and P.~Tourrenc,
\newblock {\em The Cauchy problem for the R+R**2 theories of gravity without torsion},
\newblock {\em J.\ Math.\ Phys.\ }\textbf{24} (1983) 2793, doi:10.1063/1.525659.

\bibitem{forthcoming}
P.O.~Baqui, T.~Miranda, O.F.~Piattella, F.~Sbis\`a and S.E.~Jor\'as,
\newblock {\em in preparation.}

\bibitem{Weinberg:1972}
S.~Weinberg,
\newblock {\em Gravitation and Cosmology},
\newblock John Wiley \& Sons, Inc.~(1972).

\bibitem{Chodos:1974je}
A.~Chodos, R.L.~Jaf\mbox{}fe, K.~Johnson, C.B.~Thorn and V.F.~Weisskopf,
\newblock {\em A New Extended Model of Hadrons},
\newblock {\em Phys.\ Rev.\ } \textbf{D 9} (1974) 3471--3495, doi:10.1103/PhysRevD.9.3471.

\bibitem{Peshier:1999ww}
A.~Peshier, B.~Kampfer, and G.~Sof\mbox{}f,
\newblock {\em The Equation of state of deconf\mbox{}ined matter at f\mbox{}inite chemical potential in a quasiparticle description},
\newblock {\em Phys.\ Rev.\ }\textbf{C 61} (2000) 045203, doi:10.1103/PhysRevC.61.045203, \texttt{[hep-ph/9911474]}.

\bibitem{Alford:2004pf}
M.~Alford, M.~Braby, M.W.~Paris and S.~Reddy,
\newblock {\em Hybrid stars that masquerade as neutron stars},
\newblock {\em Astrophys.\ J.\ }\textbf{629} (2005) 969--978, doi:10.1086/430902, \texttt{[nucl-th/0411016]}.

\bibitem{Schmitt:2010pn}
A.~Schmitt,
\newblock {\em Dense matter in compact stars: A pedagogical introduction},
\newblock {\em Lect.\ Notes Phys.\ }\textbf{811} (2010) 1-111, doi:10.1007/978-3-642-12866-0, \texttt{1001.3294[astro-ph]}.

\bibitem{Silbar:2003wm}
R.~R.~Silbar and S.~Reddy,
\newblock {\em Neutron stars for undergraduates},
\newblock {\em Am.\ J.\ Phys.\ }\textbf{72} (2004) 892--905
\newblock [\emph{Erratum ibid} \textbf{73} (2005) 286], doi:10.1119/1.1852544, \texttt{[nucl-th/0309041]}.

\bibitem{Douchin:2001sv}
F.~Douchin and P.~Haensel,
\newblock {\em A unif\mbox{}ied equation of state of dense matter and neutron star structure},
\newblock {\em Astron.\ Astrophys.\ }\textbf{380} (2001) 151, doi:10.1051/0004-6361:20011402, \texttt{[astro-ph/0111092]}.

\end{thebibliography}
\end{document}